\documentclass[twocolumn,aps,pre,superscriptaddress,showpacs]{revtex4-1}
\usepackage[latin9]{inputenc}
\usepackage{amsmath}
\usepackage{graphicx}
\usepackage{tikz}
\usepackage{float}

\makeatletter
 \@ifundefined{textcolor}{}
 {%
   \definecolor{BLACK}{gray}{0}
   \definecolor{WHITE}{gray}{1}
   \definecolor{RED}{rgb}{1,0,0}
   \definecolor{GREEN}{rgb}{0,1,0}
   \definecolor{BLUE}{rgb}{0,0,1}
   \definecolor{CYAN}{cmyk}{1,0,0,0}
   \definecolor{MAGENTA}{cmyk}{0,1,0,0}
   \definecolor{YELLOW}{cmyk}{0,0,1,0}
 }

\usepackage{epsfig}

\newcommand{\mb}[1]{ { \mbox{\boldmath{$#1$}}}  }

\makeatother

\begin{document}

\title{Josephson-phase-controlled interplay between correlation effects
and electron pairing in a three-terminal nanostructure}

\author{T. Doma\'{n}ski}
\email{doman@kft.umcs.lublin.pl}
\affiliation{Institute of Physics, M. Curie-Sk{\l }odowska University, PL-20-031
Lublin, Poland}

\author{M. \v{Z}onda}

\affiliation{Department of Condensed Matter Physics, Faculty of Mathematics and
Physics, Charles University in Prague, Ke Karlovu 5, CZ-121 16 Praha
2, Czech Republic}

\author{V. Pokorn\'y}

\affiliation{Institute of Physics, The Czech Academy of Sciences, Na Slovance
2, CZ-182 21 Praha 8, Czech Republic}

\author{G. G\'{o}rski}

\affiliation{Faculty of Mathematics and Natural Sciences, University of Rzesz\'{o}w,
PL-35-310 Rzesz\'{o}w, Poland}

\author{V. Jani\v{s}}

\affiliation{Institute of Physics, The Czech Academy of Sciences, Na Slovance
2, CZ-182 21 Praha 8, Czech Republic}

\author{T. Novotn\'{y}}
\email{tno@karlov.mff.cuni.cz}
\affiliation{Department of Condensed Matter Physics, Faculty of Mathematics and
Physics, Charles University in Prague, Ke Karlovu 5, CZ-121 16 Praha
2, Czech Republic}

\date{\today}
\begin{abstract}
We study the subgap spectrum of the interacting single-level quantum
dot coupled between two superconducting reservoirs, forming the Josephson-type
circuit, and additionally hybridized with a metallic normal lead.
This system allows for the phase-tunable interplay between the correlation
effects and the proximity-induced electron pairing resulting in the
singlet-doublet (0-$\pi$) crossover and the phase-dependent Kondo
effect. We investigate the spectral function, induced local pairing,
Josephson supercurrent, and Andreev conductance in a wide range of
system parameters by the numerically exact Numerical Renormalization
Group and Quantum Monte Carlo calculations along with perturbative
treatments in terms of the Coulomb repulsion and the hybridization
term. Our results address especially the correlation effects reflected
in dependencies of various quantities on the local Coulomb interaction
strength as well as on the coupling to the normal lead. We quantitatively
establish the phase-dependent Kondo temperature $\log T_{K}(\phi)\propto\cos^{2}(\phi/2)$
and show that it can be read off from the half-width of the zero-bias
enhancement in the Andreev conductance in the doublet phase, which
can be experimentally measured by the tunneling spectroscopy. 
\end{abstract}

\maketitle

\section{Introduction}

Cooper pairs of any superconducting bulk material can penetrate
a quantum impurity attached to it. This proximity-induced electron 
pairing almost completely depletes electronic states of the impurity in 
the energy regime $|\omega|\leq\Delta$ (dubbed ``subgap''), where $\Delta$ is the gap of 
the superconducting reservoir. Some in-gap ``leftovers''
can be, however, driven by the Andreev-type processes at the interface in which electrons
are converted to the Cooper pairs simultaneously emitting the holes
\cite{Balatsky-2006,*Rodero-2011}. This scattering mechanism involves
symmetrically both the particle and hole degrees of freedom, so the resulting in-gap
states appear always in pairs around the Fermi energy of the bulk superconductor. Empirical evidence for such bound (Andreev
or Yu-Shiba-Rusinov) states has been provided by STM measurements
for the magnetic (Mn, Cr) adatoms deposited on conventional (Nb, Al)
superconducting substrates \cite{Yazdani-1997,*Ji-2008,*Ji-2010,*Franke-2011}
and by numerous tunneling experiments using carbon nanotubes
\cite{Pillet-2010,*Pillet-2013,*Schindele-2014},
semiconducting nanowires \cite{Lee-2012,Lee-2014} and self-assembled
quantum dots (QDs) \cite{Deacon-2010} embedded between the normal
(N) and superconducting (S) electrodes.

\begin{figure}[t]
\includegraphics[width=1\columnwidth]{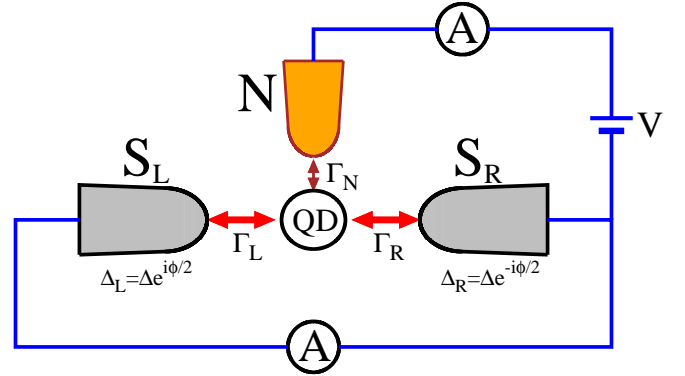} \caption{\label{fig:scheme} 
(Color online) Sketch of the three-terminal device,
where the quantum dot (QD) is coupled between the normal (N) and two
superconducting (S) leads. Phase difference $\phi$ in the Josephson
loop affects the Kondo state originating from the spin-exchange interactions
between the QD and N electrode.}
\end{figure}

In practical realizations of the N-QD-S and S-QD-S heterostructures
the impurity levels can be varied by the gate potential, changing
their even/odd occupancy \cite{Wernsdorfer-2010} which triggers evolution
of the in-gap states (affecting both their energies and spectral 
weights). Another important influence comes from the repulsive Coulomb 
interactions \cite{Bauer-2007,Yamada-2011}. In-gap states 
may cross each other (at the Fermi energy) when energies of the singly occupied 
configurations $\left|\uparrow\right>$, $\left|\downarrow\right>$ 
coincide with one of the BCS-type superpositions 
$u_{d}\left|0\right>-v_{d}\left|\uparrow\downarrow\right>$ or $v_{d}\left|0\right>
+u_{d}\left|\uparrow\downarrow\right>$ \footnote{ For $\Gamma_N\rightarrow 0$ 
and $\Delta\rightarrow\infty$ the coefficients are known explicitly \cite{Bauer-2007,Baranski-2013}:
$u_{d}^{2},v_{d}^{2}=[ 1\pm \xi_{d}/\sqrt{\xi_{d}^{2}
+(\Gamma_{L}+\Gamma_{R})^{2}/4}]/2$, with $\xi_{d} =
\varepsilon_{d}+U_{d}/2$ }.
In S-QD-S circuits this singlet-doublet quantum phase transition is
responsible for a reversal of the DC Josephson current, so called
``$0-\pi$ transition'', studied theoretically \cite{Matsuura-1977,*Glazman-1989,*Rozhkov-1999,*Yoshioka-2000,*Siano-2004,*Choi-2004,*Sellier-2005,*Novotny-2005,*Karrasch-2008,*Meng-2009,Zonda-2015,*Zonda-2016,Wentzell-2016}
and in the last decade observed experimentally \cite{vanDam-2006,*Cleuziou-2006,*Jorgensen-2007,*Eichler-2009, Maurand-2012,*Delagrange-2015,*Delagrange-2016}.

On the other hand, in N-QD-S heterostructures the singlet to doublet
crossover is smooth \cite{Zitko-2015,Jellinggaard-2016} because the 
quasiparticle states appearing in the subgap energy region
$|\omega|\leq\Delta$  acquire a finite broadening due to the hybridization with itinerant electrons
of the normal lead. This crossover has also implications on the efficiency
of the screening interactions responsible for the presence of the
Kondo effect \cite{Maurand-2013}. The recent theoretical studies
have indicated that upon approaching the crossover region from the
doublet side the antiferromagnetic spin-exchange coupling is 
substantially enhanced, amplifying the Kondo effect that can appear 
in the subgap regime solely for the spinful (doublet) configuration 
\cite{Zitko-2015,Weymann-2014,Domanski-2016}.
Although signatures of the subgap Kondo effect have been  experimentally
observed \cite{Deacon-2010,Lohneysen-2012,Chang-2013, Lee-2014} such
an intriguing evolution in the doublet-singlet crossover region is
still awaiting the systematic verification.

Inspired by Refs.~\cite{Oguri-2012,Oguri-2013,Paaske-2015} we consider
here both the Josephson and Andreev circuits arranged in the Y-shape
geometry depicted in Fig.\ \ref{fig:scheme}. This three-terminal
heterostructure would allow for a precise and phase-tunable experimental
study of the singlet-doublet transition and its influence on the Kondo
effect in the spirit of recent Refs.~\cite{Delagrange-2015,Delagrange-2016}.
Phase difference $\phi$ between the left and right superconducting
reservoirs induces DC Josephson current \textendash{} its reversal
can help to spot the $0-\pi$ crossover where the subgap quasibound
states cross each other. The role of the subgap voltage $V$ applied between
the normal electrode and Josephson circuit is twofold: It primarily
induces the Andreev current, whose differential conductance can be
used for determination of the Kondo temperature \cite{Domanski-2016},
but it may also shift the impurity level. 

\begin{figure}[t]
\includegraphics[width=1\columnwidth]{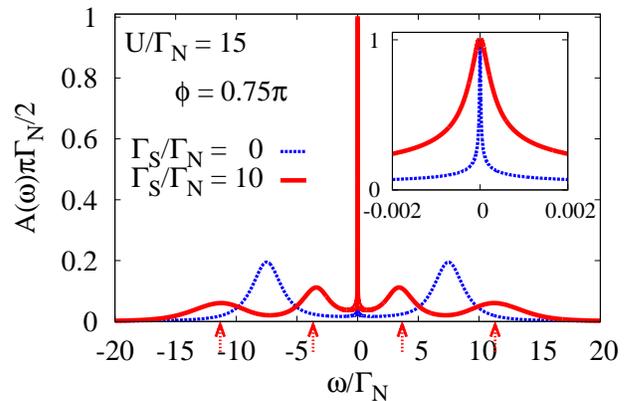}
\caption{\label{fig:spectrum_illustration} (Color online) Spectrum of the
half-filled ($\varepsilon_{d}=-U_{d}/2$) QD obtained by the NRG calculations
for $\Delta\to\infty$, $U_{d}/\Gamma_{N}=15$ assuming $\Gamma_{S}=0$
(dashed blue line) and $\Gamma_{S}/\Gamma_{N}=10$ (solid red line).
Red dashed arrows indicate the energies of Andreev bound states in
the limit $\Gamma_{N}\to0$. Inset presents details of the low-energy
(Kondo) region.}
\end{figure}

Interacting QD coupled to the usual normal Fermi liquid develops the
quasiparticle state at $\varepsilon_{d}$, its Coulomb satellite at
$\varepsilon_{d}+U_{d}$ and, at low temperatures, the Kondo peak
at zero energy (see the dashed blue line in Fig.\ \ref{fig:spectrum_illustration}).
Broadening of the charge quasiparticle states is controlled by $\Gamma_{N}$,
whereas the zero-energy peak depends on the effective exchange
potential \cite{S-W}. Its half-width can be related to the characteristic
Kondo scale $k_{B}T_{K}$ \cite{Haldane-78,*Tsvelik-83}. In our setup
involving superconducting leads (Fig.\ \ref{fig:scheme}) the QD
spectrum is completely different. We illustrate the qualitative changes
by the solid line in Fig.\ \ref{fig:spectrum_illustration}, obtained
for the doublet ground state configuration at the particle-hole symmetric
point ($\varepsilon_{d}=-U_{d}/2$) in the limit $\Delta\to\infty$.
Instead of the charge peaks at $\pm U_{d}/2$ of the normal case we
observe four Andreev quasibound states at $\pm[U_{d}/2\pm\Gamma_{S}\cos(\phi/2)]$
\cite{Baranski-2013} coexisting with the zero-energy Kondo peak,
which has, however, a substantially broadened width in comparison to the
$\Gamma_{S}=0$ value \cite{Zitko-2015,Domanski-2016}. These effects
are caused by interplay of the proximity effect induced by the couplings
$\Gamma_{L,R}$ with the correlations due to $U_{d}$, and they depend
on the coupling $\Gamma_{N}$ to the normal electrode \cite{Zitko-2015}. 

The subtle influence of the normal electrode has already been studied
by K\"onig and coworkers \cite{Governale-2008,*Futterer-2013} by means
of the perturbative real-time diagrammatic expansion with respect
to the hybridization terms. Their analysis has been next extended
to the Kondo regime using the numerical renormalization group (NRG)
calculations in the infinite gap $\Delta\to\infty$ limit \cite{Oguri-2012,Oguri-2013}
and/or for finite gap but just a single superconducting reservoir
(i.e., no option of tuning the phase difference $\phi$) \cite{Zitko-2015}.
Additional information has been recently obtained from the T-matrix
studies of the classical and quantum spin impurity \cite{Paaske-2015},
revealing the phase $\phi$ dependence of the subgap states \textemdash{}
however, the employed classical-spin and perturbative quantum approximations
do not address the Kondo effect.

Our work extends the previous studies of the subgap states \cite{Governale-2008,*Futterer-2013,Oguri-2013,Paaske-2015},
focusing on the phase-tunable Kondo physics and its possible empirical
observability. In what follows we explore the phase dependence of
the Kondo effect, using: (i) perturbative treatment of the hybridization
$V_{{\bf k}N}$ by the Schrieffer-Wolff transformation \cite{S-W,Bravyi-2011,Zamani-2016}
generalized to the superconducting Hamiltonian \cite{Domanski-2016},
(ii) the self-consistent second order perturbative treatment of the
Coulomb potential $U_{d}$ \cite{Zonda-2015,*Zonda-2016}, (iii) the
NRG \cite{Bulla-2008,Zitko-2015,Zitko-2016}, and QMC \cite{Gull-2011,Luitz-2010,*Luitz-2012}
calculations. We determine the Kondo temperature for a wide range
of parameters by a suitable combination of these methods and obtain
its universal phase scaling $\mbox{{\rm ln}}T_{K}(\phi)\propto\mbox{{\rm cos}}^{2}\left(\phi/2\right)$.

The paper is organized as follows. In Sec.~\ref{sec:methods}, after
introducing the model Hamiltonian in Sec.~\ref{subsec:micro-model},
we present the four used theoretical methods \textemdash{} namely,
the effective spin-exchange interaction via the generalized Schrieffer-Wolff
transformation in Sec.~\ref{subsec:sw}, second-order perturbation
theory in $U_{d}$ in Sec.~\ref{subsec:perturb}, NRG in Sec.~\ref{subsec:NRG},
and QMC in Sec.~\ref{subsec:CT-HYB}. We then analyze the obtained
results for the spectral function (Sec.~\ref{sec:Spectral-function}),
Josephson current (Sec.~\ref{sec:Josephson-current}), and finally
for the Andreev conductance in Sec.~\ref{sec:Andreev-conductance}.
We summarize our study in Sec.~\ref{sec:Summary}.

\section{Model and methods\label{sec:methods}}

\subsection{Microscopic formulation\label{subsec:micro-model}}
Three-terminal system with the quantum dot depicted in Fig.\ \ref{fig:scheme}
can be described by the Anderson model Hamiltonian 
\begin{subequations}\label{eq:model}
\begin{eqnarray}
\hat{H} & = & \sum_{\sigma}\varepsilon_{d}\hat{d}_{\sigma}^{\dagger}\hat{d}_{\sigma}+U_{d}\hat{n}_{d\uparrow}\hat{n}_{d\downarrow}+\sum_{\beta}\hat{H}_{\beta}\\
 & + & \sum_{{\bf k},\sigma}\sum_{{\beta}}\left(V_{{\bf k}\beta}\;\hat{d}_{\sigma}^{\dagger}\hat{c}_{{\bf k}\sigma\beta}+V_{{\bf k}\beta}^{*}\hat{c}_{{\bf k}\sigma\beta}^{\dagger}\hat{d}_{\sigma}\right),\nonumber 
\end{eqnarray}
where the reservoirs $\beta=N,L,R$ of itinerant electrons are given
by 
\begin{eqnarray}
\hat{H}_{N} & = & \sum_{{\bf k},\sigma}\xi_{{\bf k}N}\hat{c}_{{\bf k}\sigma N}^{\dagger}\hat{c}_{{\bf k}\sigma N},\\
\hat{H}_{S} & = & \sum_{{\bf k},\sigma}\xi_{{\bf k}S}\hat{c}_{{\bf k}\sigma S}^{\dagger}\hat{c}_{{\bf k}\sigma S}\\
 & - & \Delta\sum_{{\bf k}}\left(e^{i\phi_{S}}\hat{c}_{{\bf k}\uparrow S}^{\dagger}\hat{c}_{-{\bf k}\downarrow S}^{\dagger}+e^{-i\phi_{S}}\hat{c}_{-{\bf k}\downarrow S}\hat{c}_{{\bf k}\uparrow S}\right)\nonumber 
\end{eqnarray}
\end{subequations}
and $S=L,R$. We use the second quantization notation, in which $\hat{d}_{\sigma}^{(\dag)}$
annihilates (creates) an electron with spin $\sigma$ at the QD energy
level $\varepsilon_{d}$. As usually $\hat{n}_{d\sigma}=\hat{d}_{\sigma}^{\dagger}\hat{d}_{\sigma}$
stands for the number operator and $U_{d}$ is the repulsive Coulomb
potential between opposite spin electrons. The QD is coupled to the
mobile electrons of $\beta$ reservoir by the matrix elements $V_{{\bf k}\beta}$
and the energies $\xi_{{\bf k}\beta}=\varepsilon_{{\bf k}}-\mu_{\beta}$
are measured with respect to the chemical potentials $\mu_{\beta}$
($\mu_{L}=\mu_{R}=0$), which can be detuned by the voltage $\mu_{N}=eV$.
We assume that both superconductors are isotropic and are characterized
by the complex order parameters $\Delta e^{i\phi_{S}}$.

Our major concern here are the states appearing in the low-energy
subgap regime $|\omega|\leq\Delta$ therefore we can assume the couplings
$\Gamma_{\beta}=2\pi\sum_{{\bf k}}|V_{{\bf k}\beta}|^{2}\;\delta(\omega\!-\!\xi_{{\bf k}\beta})$
to be constant. In realistic systems the Coulomb potential by far
exceeds the superconducting gap $U_{d}\gg\Delta$ and, consequently,
the outer Andreev states are pushed towards the band edges or even
into the continuum \cite{Rodero-2011,Paaske-2015}. For this reason
we shall disregard them from our considerations, which are primarily
focused on the deep subgap regime. To simplify the formulas we focus
on the symmetric couplings $\Gamma_{L}=\Gamma_{R}\equiv\Gamma_{S}/2$
of the Josephson loop which is phase biased by $\phi\equiv\phi_{L}-\phi_{R}$.
Asymmetric coupling $\Gamma_{L}\neq\Gamma_{R}$ can be handled easily
by the replacement $\cos(\phi/2)\to\sqrt{1-4\Gamma_{L}\Gamma_{R}\sin^{2}(\phi/2)/(\Gamma_{L}+\Gamma_{R})^{2}}$
\cite{Kadlecova-2016}. Our numerical calculations are mostly performed
for the half-filled QD ($\varepsilon_{d}=-U_{d}/2$). In what follows
we address: (i) emergence of the subgap Kondo state in the limit $k_{B}T_{K}\ll\Delta$,
(ii) its evolution upon approaching the doublet to singlet crossover,
and (iii) observable signatures of the phase-dependent Kondo effect
in the transport properties.

\subsection{Effective spin exchange interaction\label{subsec:sw}}

To quantify analytically the phase dependence of the Kondo features
we estimate the effective spin-exchange potential driven by the hybridization
term $\hat{V}_{N}=\sum_{{\bf k},\sigma}\left(V_{{\bf k}N}\;\hat{d}_{\sigma}^{\dagger}\hat{c}_{{\bf k}\sigma N}+V_{{\bf k}N}^{*}\hat{c}_{{\bf k}\sigma N}^{\dagger}\hat{d}_{\sigma}\right)$
and the local Coulomb interaction $U_{d}\hat{n}_{d\uparrow}\hat{n}_{d\downarrow}$.
This can be done rather straightforwardly in the SC atomic limit of
very large gap $\Delta\to\infty$ by adopting the Schrieffer-Wolff
approach \cite{S-W,Bravyi-2011,Zamani-2016} to the effective Hamiltonian
\cite{Tanaka-2007-jpsj,Karrasch-2008,Rodero-2011,Oguri-2013} 
\begin{eqnarray}
\hat{H_{\mathrm{eff}}} & = & \hat{H}_{N}+\hat{V}_{N}+\sum_{\sigma}\varepsilon_{d}\hat{d}_{\sigma}^{\dagger}\hat{d}_{\sigma}+U_{d}\hat{n}_{d\uparrow}\hat{n}_{d\downarrow}\nonumber \\
 & - & \Delta_{d}(\phi)\left(\hat{d}_{\uparrow}^{\dagger}\hat{d}_{\downarrow}^{\dagger}+\hat{d}_{\downarrow}\hat{d}_{\uparrow}\right).\label{proximized}
\end{eqnarray}
with the real induced pairing potential $\Delta_{d}(\phi)\equiv\Gamma_{S}\cos(\phi/2).$
Apart from the explicit expression for $\Delta_{d}$ in terms of model
parameters this Hamiltonian is identical to that of Ref.~\cite{Domanski-2016},
Eq.~(2). We can therefore follow exactly the same extended Schrieffer-Wolff
(SW) procedure eliminating perturbatively the hybridization term $\hat{V}_{N}$
from the Hamiltonian \eqref{proximized} by a canonical unitary transformation
to obtain the spin-exchange interaction 
\begin{eqnarray}
\hat{H}_{\mathrm{exch}}=-\sum_{{\bf k},{\bf p}}J_{{\bf k,p}}(\phi)\;\;\hat{{\bf S}}_{d}\cdot\hat{{\bf S}}_{{\bf k}{\bf p}}\label{exch_term}
\end{eqnarray}
between the QD spin $\hat{{\bf S}}_{d}$ and the spins of the normal
lead electrons $\hat{{\bf S}}_{{\bf k}{\bf p}}$. The coupling potential
near the Fermi energy $J_{\mathrm{{\bf k_{F},}\mathbf{k_{F}}}}(\phi)$
is given by (compare with Eq.~(19) of Ref.~\cite{Domanski-2016})
\begin{eqnarray}
J_{{\bf k}_{F},{\bf k}_{F}}(\phi)=\frac{U_{d}\left|V_{{\bf k}_{F}}\right|^{2}}{\varepsilon_{d}\left(\varepsilon_{d}+U_{d}\right)+\Delta_{d}^{2}(\phi)}\label{generalized_SW}
\end{eqnarray}
 with $\phi$-dependent $\Delta_{d}(\phi)$ in the present case. The
vanishing denominator of Eq.\ \eqref{generalized_SW} $\varepsilon_{d}\left(\varepsilon_{d}+U_{d}\right)+\Delta_{d}^{2}(\phi_{c})=0$
defines the singlet-doublet transition point (if it exists for any
$\phi_{c}\in(0,2\pi)$) \cite{Oguri-2013}
\begin{equation}
\phi_{c}=2\arccos\frac{\sqrt{|\varepsilon_{d}|(\varepsilon_{d}+U_{d})}}{\Gamma_{S}}.\label{ala_Bauer}
\end{equation}
The ferromagnetic regime can be, however, completely disregarded,
because it would have no effect on the BCS-type (spinless) configuration
$u_{d}\left|0\right>-v_{d}\left|\downarrow\uparrow\right>$.

We can determine the effective Kondo temperature for the doublet (spinful)
configuration using the well-known formula \cite{Haldane-78,*Tsvelik-83}
$T_{K}\propto\mbox{{\rm exp}}\left\{ -1/\left[2\rho(\varepsilon_{F})J_{{\bf k}_{F}{\bf k}_{F}}\right]\right\} $,
where $\rho(\varepsilon_{F}$) is the density of states of the normal
lead at the Fermi level. We thus obtain 
\begin{eqnarray}
T_{K}(\phi)=\eta\;\frac{\sqrt{\Gamma_{N}U_{d}}}{2}\;{\rm exp}\!\left[\pi\;\frac{\varepsilon_{d}\left(\varepsilon_{d}+U_{d}\right)+\Delta_{d}^{2}(\phi)}{\Gamma_{N}U_{d}}\right]\label{T_K}
\end{eqnarray}
with $\eta$ being of the order of unity (its specific value depending
on the exact definition of $T_{K}$). Its comparison to the normal
case corresponding to $\Gamma_{S}=0$ and hence $\Delta_{d}(\phi)=0$
yields the following phase scaling of the relative Kondo temperature
\begin{equation}
\ln\left(\frac{T_{K}}{T_{K}^{N}}\right)=\frac{\pi\;\Gamma_{S}^{2}}{\Gamma_{N}U_{d}}\;\cos^{2}\left(\frac{\phi}{2}\right)\label{effective_T_K}
\end{equation}
valid in the doublet region $\cos^{2}({\phi}/{2})<|\varepsilon_{d}|(U_{d}+\varepsilon_{d})/\Gamma_{S}^{2}$
(equal to $(U_{d}/2\Gamma_{S})^{2}$ for the particle-hole symmetric
case). 

In experiments the value of the SC gap $\Delta$ is typically smaller
or comparable to the other parameters such as $\Gamma_{S}$ and $U_{d}$.
Thus, the above derivation cannot be directly used. Extension of the
above Schrieffer-Wolff procedure to the finite-$\Delta$ case remains
an open question, which is further complicated by the lack of the
numerical support by NRG, which practically cannot handle the associated
three-channel setup (see below in Sec.~\ref{subsec:NRG}). We wish
to point out explicitly that the extension of the Schrieffer-Wolff
procedure to the superconducting case by Salomaa \cite{Salomaa-1988}
and its further application to the NS case \cite{Zitko-2015} based
on the perturbative treatment of \emph{both} $\Gamma_{N}$ and $\Gamma_{S}$
is not sufficient for our present purpose of determining the phase-dependent
change of the Kondo temperature. Due to its perturbative nature in
$\Gamma_{S}$ it fails completely to capture the effect of the SC
lead(s) on the exchange coupling between the localized spin and the
\emph{normal lead} electrons, cf.~Eq.~(B19) in Ref.~\cite{Zitko-2015}.
Our result \eqref{generalized_SW}, on the contrary, contains the
effects of the SC leads in a fully resummed fashion via the $\Delta_{d}^{2}(\phi)$
term in the denominator. It is not currently clear to us how to achieve
such a nonperturbative result in the finite-$\Delta$ case, where
the SC leads cannot be so straightforwardly incorporated into the
effective Hamiltonian \eqref{proximized}. 

\subsection{Perturbative treatment in $U_{d}$\label{subsec:perturb}}

It has been recently shown for the S-QD-S \cite{Zonda-2015,*Zonda-2016}
and N-QD-S \cite{Domanski-2016} junctions, that some of the NRG results
can be reproduced using the self-consistent perturbative treatment
of the interaction $U_{d}\hat{n}_{d\downarrow}\hat{n}_{d\uparrow}$.
This fact encouraged us to apply the same technique to our three-terminal
system in Fig.\ \ref{fig:scheme}. For treating the mixed particle
and hole degrees of freedom we have to formulate the diagrammatic
expansions for the (real-time) matrix Green's function
${\mb G}(t,t')\!=\!\langle\!\langle\hat{\Psi}_{d}(t);\hat{\Psi}_{d}^{\dagger}(t')\rangle\!\rangle$
in the Nambu representation, where $\hat{\Psi}_{d}\equiv(\hat{d}_{\uparrow},\hat{d}_{\downarrow}^{\dagger})^{T}$.
Its Fourier transform obeys the Dyson equation 
\begin{eqnarray}
{\mb G}^{-1}(\omega)=\left(\begin{array}{cc}
\omega\!-\!\varepsilon_{d} & 0\\
0 & \omega\!+\!\varepsilon_{d}
\end{array}\right)-{\mb\Sigma}(\omega),\label{GF}
\end{eqnarray}
where the self-energy ${\mb\Sigma}(\omega)$ accounts for the three
external reservoirs and for the correlation effects due to $U_{d}$.
In the absence of the correlations the self-energy $\mb{\Sigma}^{0}(\omega)\equiv\left.\mb{\Sigma}(\omega)\right|_{U_{d}=0}$
can be obtained exactly. In the wide band limit it is given by the
following formula \cite{Bauer-2007,Yamada-2011} 
\begin{eqnarray}
\mb{\Sigma}^{0}(\omega) & = & -\;\frac{i\Gamma_{N}}{2}\;\left(\begin{array}{cc}
1 & 0\\
0 & 1
\end{array}\right)-\Gamma_{S}\mbox{{\rm cos}}\left(\frac{\phi}{2}\right)\left(\begin{array}{cc}
1 & \frac{\Delta}{\omega}\\
\frac{\Delta}{\omega} & 1
\end{array}\right)\nonumber \\
 & \times & \left\{ \begin{array}{ll}
\frac{\omega}{\sqrt{\Delta^{2}-\omega^{2}}} & \mbox{{\rm for }}|\omega|<\Delta\\
\frac{i\;|\omega|}{\sqrt{\omega^{2}-\Delta^{2}}} & \mbox{{\rm for }}|\omega|>\Delta
\end{array}\right..\label{selfenergy_0}
\end{eqnarray}
Equation (\ref{selfenergy_0}) describes the induced on-dot pairing via
the off-diagonal term proportional to $\Gamma_{S}$ and the finite
life-time effects which enter through the imaginary parts of the self-energy
$\mb{\Sigma}^{0}(\omega)$. 
Extending our previous studies \cite{Domanski-2016,Zonda-2015,*Zonda-2016}
to the present three-terminal heterostructure (Fig.\ \ref{fig:scheme})
we express the full $\mb{\Sigma}(\omega)$ using the second-order
perturbation theory (SOPT) \cite{Zonda-2015,*Zonda-2016,Vecino-2003} 
\begin{subequations}\label{sigma}
\begin{eqnarray}
{\mb\Sigma}_{11}(\omega) & = & {\mb\Sigma}_{11}^{0}(\omega)%-i\frac{\Gamma_{N}}{2}
+U_{d}\langle\hat{d}_{\downarrow}^{\dagger}\hat{d}_{\downarrow}\rangle\label{diag_sigma}\\
 & + & U_{d}^{2}\int\limits _{-\infty}^{\infty}{\frac{-\frac{1}{\pi}\mbox{{\rm Im}}{{\mb\Sigma}_{11}^{(2)}(\omega')}}{{\omega-\omega'+i0^{+}}}d\omega'},\nonumber \\
{\mb\Sigma}_{12}(\omega) & = & {\mb\Sigma}_{12}^{0}(\omega)%-\Gamma_{S}\;\cos{\left(\frac{\phi}{2}\right)}
+U_{d}\langle\hat{d}_{\downarrow}\hat{d}_{\uparrow}\rangle\label{offdiag_sigma}\\
 & - & U_{d}^{2}\int\limits _{-\infty}^{\infty}{\frac{-\frac{1}{\pi}\mbox{{\rm Im}}{{\mb\Sigma}_{12}^{(2)}(\omega')}}{{\omega-\omega'+i0^{+}}}d\omega'}.\nonumber 
\end{eqnarray}
The imaginary parts of the second-order contributions
${\mb\Sigma}_{ij}^{(2)}(\omega)$ are expressed by the following convolutions
\begin{eqnarray}
-\frac{1}{\pi}\mbox{{\rm Im}}{\mb\Sigma}_{11(12)}^{(2)}(\omega) & = & \int\limits _{-\infty}^{\infty}\left[{\mb\Pi}_{1}(\omega+\omega')\rho_{22(12)}^{+}(\omega')\right.\nonumber \\
 & + & \left.{\mb\Pi}_{2}(\omega+\omega')\rho_{22(12)}^{-}(\omega')\right]\;d\omega',\label{imag_part}\\
{\mb\Pi}_{1(2)}(\omega) & = & \int\limits _{-\infty}^{\infty}\left[\rho_{11}^{-(+)}(\omega')\rho_{22}^{-(+)}(\omega-\omega')\right.\nonumber \\
 & - & \left.\rho_{12}^{-(+)}(\omega')\rho_{21}^{-(+)}(\omega-\omega')\right]\;d\omega',\label{Pi_prop}
\end{eqnarray}
\end{subequations}
where the auxiliary functions $\rho_{ij}^{\pm}(\omega)\equiv-\mathrm{Im}\,{\mb G}_{ij}^{\mathrm HF}(\omega\!+\!i0^{+})f(\pm\omega)/\pi$
(with the standard Fermi-Dirac distribution $f(\omega)\equiv1/\left[1+\mbox{{\rm exp}}\left(\omega/k_{B}T\right)\right]$)
denote the occupancies obtained at the Hartree-Fock approximation
level $\sum_{\sigma}\varepsilon_{d}\hat{d}_{\sigma}^{\dagger}\hat{d}_{\sigma}+U_{d}\;\hat{n}_{d\uparrow}\hat{n}_{d\downarrow}\rightarrow\sum\limits _{\sigma}{\left({\varepsilon_{d}+U_{d}\langle\hat{n}_{d-\sigma}\rangle}\right)\hat{n}_{d\sigma}}+U_{d}(\langle\hat{d}_{\downarrow}\hat{d}_{\uparrow}\rangle\hat{d}_{\uparrow}^{\dagger}\hat{d}_{\downarrow}^{\dagger}+\mbox{{\rm h.c.}})$.

\subsection{NRG\label{subsec:NRG}}

The most reliable (unbiased) information about the relationship between
the electron correlations and induced pairing can be obtained within
a nonperturbative scheme of the numerical renormalization group (NRG)
technique \cite{Bulla-2008}. We have performed NRG calculations using
the open source code NRG LJUBLJANA \cite{ZitkoPruschke-2009,*Ljubljana-code}.
For the infinite SC gap $\Delta\to\infty$ the NRG reduces to a single
channel problem \cite{Oguri-2013,Tanaka-2007}, where we have set
the logarithmic discretization parameter $\Lambda=1.9$. The case
of finite $\Delta$ is in general a three-channel problem and trustworthy
NRG calculations for the Anderson model with three channels are extremely
demanding. Fortunately, in the special case of $\phi=0$ one can effectively
reduce this problem to two channels. The NRG data for this case have
been calculated with $\Lambda=4$, as is common for such double-channel
models.

\subsection{CT-HYB\label{subsec:CT-HYB}}

We have also utilized continuous-time quantum Monte-Carlo (CT-QMC)
calculations~\cite{Gull-2011} in order to study the effect of finite
temperature on the system and to obtain results in regions, where
NRG is ineffective. Since Hamiltonian \eqref{eq:model} does not conserve
the electron number, we perform the canonical particle-hole transformation
\begin{equation}
\hat{d}_{\uparrow}^{\dag}\rightarrow\tilde{d}_{\uparrow}^{\dag},\quad\hat{d}_{\downarrow}^{\dag}\rightarrow\tilde{d}_{\downarrow}^{\phantom{\dag}},\quad\hat{c}_{\mathbf{k},\uparrow,\beta}^{\dag}\rightarrow\tilde{c}_{\mathbf{k},\uparrow,\beta}^{\dag},\quad\hat{c}_{\mathbf{k},\downarrow,\beta}^{\dag}\rightarrow\tilde{c}_{\mathbf{-k},\downarrow,\beta}^{\phantom{\dag}}
\end{equation}
which was already used by Luitz and Assaad~\cite{Luitz-2010} to
include superconductivity in the CT-QMC calculations. The new quasiparticles
described by $\tilde{c}$ and $\tilde{d}$ operators are identical
to electrons in the spin-up sector and to holes in the spin-down sector.
This transformation maps our system to a one-band Anderson model with
attractive interaction $-U_{d}$ and local energy term $\varepsilon_{d}(\tilde{n}_{\uparrow}-\tilde{n}_{\downarrow})$,
where $\tilde{n}_{\sigma}=\tilde{d}_{\sigma}^{\dag}\tilde{d}_{\sigma}^{\phantom{\dag}}$.

We use the \textsc{TRIQS} hybridization-expansion continuous-time
(CT-HYB) quantum Monte Carlo solver~\cite{Seth-2016} based on the
\textsc{TRIQS} libraries~\cite{Parcollet-2015}. We consider a flat
band in the leads of finite width $D=20\Gamma_{S}$. The induced gap
$\Delta_{I}$ and Josephson current $I_{J}$ are obtained from the
anomalous part of the imaginary-time Green's function \cite{Luitz-2010,*Luitz-2012}.

\section{Spectral function and the Kondo scale \label{sec:Spectral-function}}

\begin{figure}[t]
\includegraphics[width=1\columnwidth]{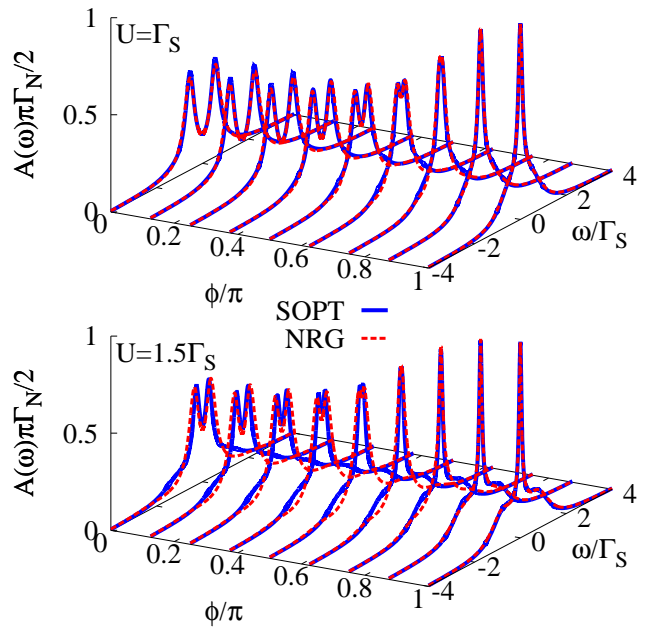} \caption{\label{fig:spectral_function} (Color online) Phase dependence of
the diagonal spectral function $A(\omega)$ of the half-filled QD
in the large-gap limit $\Delta\to\infty$ obtained by SOPT (blue solid
lines) and NRG calculations (red dashed curves). Results are presented
for $\Gamma_{N}/\Gamma_{S}=0.5$ and two values of the Coulomb potential
$U_{d}/\Gamma_{S}=1$ (top panel) and $U_{d}/\Gamma_{S}=1.5$ (bottom
panel).}
\end{figure}

Phase evolution of the (diagonal) spectral function $A(\omega)\equiv-\mbox{{\rm Im}}{\mb G}_{11}(\omega+i0^{+})/\pi$
obtained by the second order perturbation theory (SOPT; Eqs. \eqref{sigma})
and compared with the NRG results is displayed in Fig.\ \ref{fig:spectral_function}.
The panels correspond to the half-filled QD case ($\varepsilon_{d}=-U_{d}/2$)
in the large-gap regime $\Delta\to\infty$ with $\Gamma_{N}=\Gamma_{S}/2$
and different Coulomb potentials, as indicated. In both cases we see
an excellent agreement between the SOPT and NRG data with some quantitative
differences noticeable only in the lower panel for intermediate $\phi$'s.

For both values of the Coulomb potential $U_{d}$, the subgap electronic
spectrum shows either the absence (for the spinless BCS-type configuration,
close to the $0$ phase for small $\phi$'s) or presence (for the
spinful doublet state, close to $\pi$ phase for $\phi\gtrsim\phi_{c}=2\arccos\left(U_{d}/2\Gamma_{S}\right)$
given by Eq.~\eqref{ala_Bauer}) of the central Kondo-like peak around
zero frequency. Such a phase evolution we assign to the efficiency
or lack of the screening interactions. 

\begin{figure}[htpb]
\includegraphics[width=1\columnwidth]{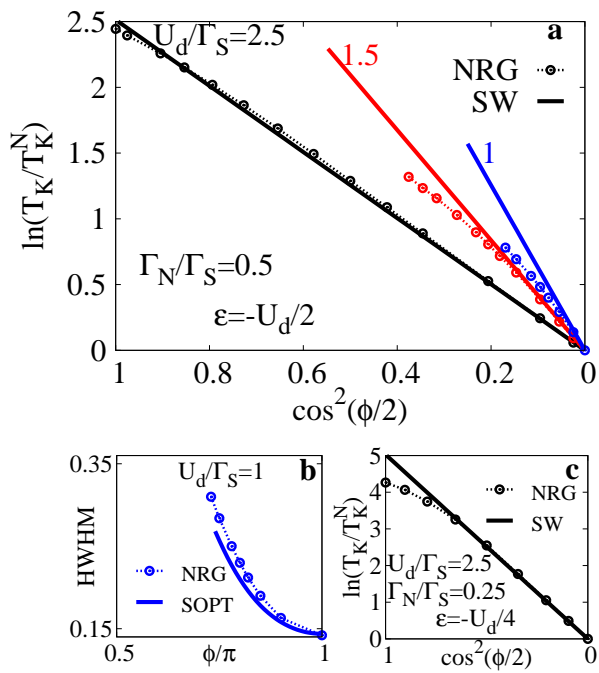} \caption{\label{fig:T_K_NRG} (Color online) (\textbf{a}) Phase dependence
of the Kondo temperature $T_{K}$ obtained by NRG for $\Gamma_{N}/\Gamma_{S}=0.5$
and several potentials $U_{d}/\Gamma_{S}$ as indicated (dotted lines
with bullets) and compared to the generalized Schrieffer-Wolff prediction
\eqref{effective_T_K} (full lines) ending at the transition point
determined by Eq.~\eqref{ala_Bauer}. (\textbf{b}) Comparison of
the absolute HWHM from the NRG and SOPT for relatively small $U_{d}=\Gamma_{S}$.
(\textbf{c}) Analogously to panel \textbf{a} for the large $U_{d}=2.5\Gamma_{S}$
and smaller normal coupling $\Gamma_{N}=\Gamma_{S}/4$ but away from
the half-filling ($\varepsilon_{d}=-U_{d}/4$). In all panels we use
the large-gap limit $\Delta\to\infty$ and in panels \textbf{a} and
\textbf{c} the $T_{K}$ is normalized to $T_{K}^{N}$ corresponding
to $\Gamma_{S}=0$ (i.e., in the absence of the proximity effect).}
\end{figure}

Figure \ref{fig:T_K_NRG} shows the phase-dependent Kondo temperature
determined as the half-width at half maximum (HWHM) of the zero-energy
peak by the NRG and SOPT and compared to the generalized Schrieffer-Wolff
prediction \eqref{effective_T_K}. We observe a very good correspondence
between NRG and SOPT for small enough $U_{d}$ (panel \textbf{b})
and between NRG and SW for larger values of $U_{d}$ (panels \textbf{a}
and \textbf{c}). The deviations observed between the NRG data and
SW prediction occur close to the $0-\pi$ crossover where the central
Kondo-like peak merges with the broadened Andreev states and the identification
of the Kondo scale from the HWHM is no longer reliable. Furthermore,
the effective exchange coupling \eqref{generalized_SW} diverges at
the $0-\pi$ transition where the perturbative SW approach necessarily
breaks down and, thus, the SW prediction close to the transition stops
working. Nonperturbative treatment beyond the simple SW approach
based on the flow equations (corrected version of Ref.~\cite{Zapalska-2014})
shows sharp but smooth features at the transition point; this issue
is, however, beyond our current scope and the refined results will
be published elsewhere. For the largest value of $U_{d}$ in the particle-hole
symmetric case (black curves in panel \textbf{a}) when the system
is close to the $\pi$ phase for all $\phi$'s the agreement between
the two methods is nearly perfect. 

In general, the trend extracted from all the available data confirms
the enhancement of $T_{K}$ (i.e., broadening of the Kondo-like peak
with respect to the normal case as seen already in the inset of Fig.~\ref{fig:spectrum_illustration})
upon approaching the singlet state from the doublet side. For sufficiently
large values of $U_{d}$ the Kondo scale enhancement follows quite
well the analytical formula \eqref{effective_T_K} above $\phi_{c}$
\eqref{ala_Bauer}. The logarithm of the enhancement is proportional
to $\cos^{2}(\phi/2)$ with the proportionality constant inversely
decaying with increasing $U_{d}$ \textemdash{} for the strong Coulomb
interaction $U_{d}$ the relative value of the Kondo scale hardly
changes with respect to $\phi$ because the correlations dominate
over the induced electron pairing. For the opposite case of relatively
small $U_{d}$, the Kondo state survives only when the Josephson phase
$\phi$ is nearby $\pi$ (i.e., $\phi\gtrsim\phi_{c}$), where the
quantum interference between superconducting reservoirs is destructive. 

\begin{figure}[t]
\includegraphics[width=1\columnwidth]{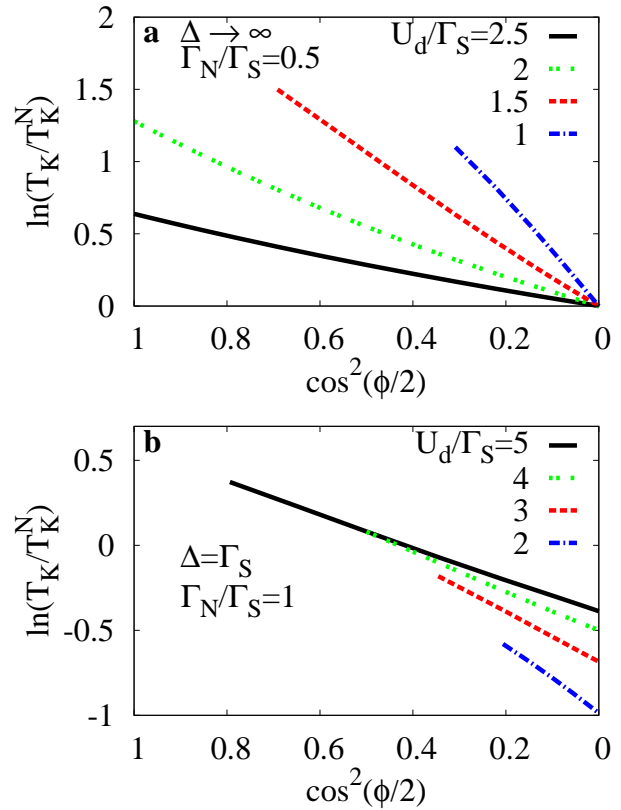} \caption{\label{fig:T_K_SOPT} (Color online) Universal phase dependence $\ln T_{K}\propto\cos^{2}(\phi/2)$
of the normalized Kondo temperature $T_{K}$ obtained by the SOPT
at half-filling for several $U_{d}/\Gamma_{N}$ values and large (\textbf{a})
and finite (\textbf{b}) gap. }
\end{figure}

The above conclusions are based on the results in the large-gap limit
$\Delta\to\infty$ where the NRG works reliably. However, as already
mentioned this limit is not experimentally realistic and one may wonder
whether these statements do survive also for finite $\Delta$ cases.
Unfortunately, the case of simultaneously finite $\Delta$ and $\phi$
is practically very difficult to handle by NRG and we must thus resort
to an approximative treatment based on SOPT. As seen in Fig.~\ref{fig:T_K_NRG},
panel \textbf{b} SOPT and NRG yield comparable results and this encourages
us to address the finite-$\Delta$ case by SOPT with results summarized
in Fig.~\ref{fig:T_K_SOPT}. The two panels describe (\textbf{a})
the large-gap $\Delta\to\infty$ case corresponding to the panel \textbf{a}
of Fig.~\ref{fig:T_K_NRG} and (\textbf{b}) fairly realistic finite
$\Delta=\Gamma_{S}$ case at half-filling for the same set of $U_{d}/\Gamma_{N}$
ratios as in panel \textbf{a}. In both cases we do observe the universal
linear behavior $\ln T_{K}\propto\cos^{2}(\phi/2)$ of the logarithm
of the Kondo scale. 

Nevertheless, it is at place to mention several problematic aspects
of this finding here. First, although the logarithm of the Kondo scale
is proportional to $\cos^{2}(\phi/2)$ and the proportionality coefficient
correctly decreases with increasing $U_{d}$, its value is not captured
quantitatively precisely as can be seen by comparing the black lines
(corresponding to $U_{d}/\Gamma_{S}=2.5$) in panels \textbf{a} of
Figs.~\ref{fig:T_K_NRG} and \ref{fig:T_K_SOPT} \textemdash{} they
should be identical but they are clearly not. This is not so surprising
as this case corresponds to a rather large interaction value $U_{d}=5\Gamma_{N}$
where the SOPT, being a simple perturbation scheme, is not expected
to yield very reliable results. For the smaller values of $U_{d}$
our SOPT findings should be even quantitatively correct as shown in
panel \textbf{b} of Fig.~\ref{fig:T_K_NRG}. However, for such small
values of the interaction, the central peak in the spectral function
cannot be attributed to a true Kondo effect, rather to its weak coupling
precursor. Despite all these objections we still believe that the
qualitative conclusions drawn from the SOPT results, which confirm
the universal phase dependence of the Kondo scale even for finite
SC gap, are valid. 

\begin{figure}
\includegraphics[width=1\columnwidth]{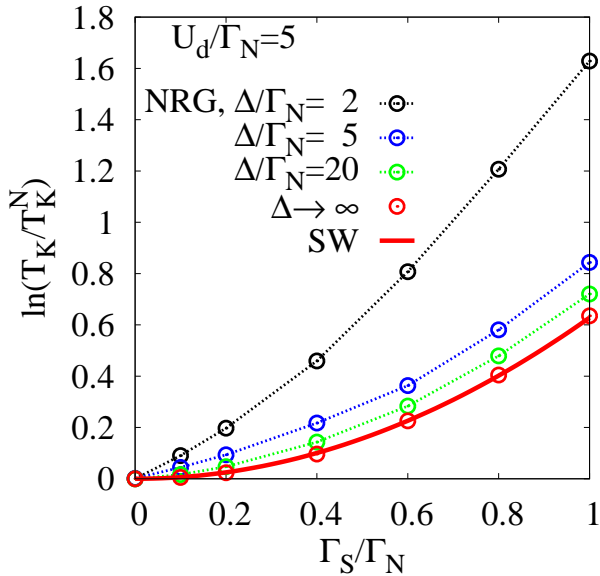}\caption{\label{fig:Kondo_finite_gap}(Color online) SC-induced enhancement
of the Kondo temperature at $\phi=0$ calculated by NRG as a function
of $\Gamma_{S}$ for several finite values of the gap $\Delta$. The
bottom curve corresponds to the infinite-gap limit and is perfectly
determined by the SW formula \eqref{effective_T_K}.}
\end{figure}

Finally, we address the issue mentioned at the end of the Schrieffer-Wolff
section \ref{subsec:sw}, namely, the effect of finite value of
the SC gap $\Delta$ on the enhancement of the Kondo scale and its
relation to the infinite-$\Delta$ limit. To obtain quantitatively
reliable results we resort to the NRG calculations, which can be for
finite $\Delta$ performed only for $\phi=0$ (see above). We plot
the SC-enhanced Kondo temperature as a function of the coupling to
the superconducting leads $\Gamma_{S}$ for several values of $\Delta$
in Fig.~\ref{fig:Kondo_finite_gap}. The parameter values are chosen
such that there is no $0-\pi$ transition in the whole range of $\phi$'s
and the system is close to the $\pi$ phase even for $\phi=0$, where
we evaluate the spectral function and read off the Kondo scale. We
see that the enhancement gets even more pronounced with decreasing
$\Delta$ (within reasonable limits \textemdash{} we always work in
the regime where the Kondo peak is well inside the SC gap, i.e., $T_{K}\ll\Delta$).
The values of $\Delta$ comparable to $\Gamma_{S}$ (corresponding
to the topmost curve in Fig.~\ref{fig:Kondo_finite_gap}) are experimentally
fully justified, but we have no analytical theory for them, unlike
in the $\Delta\to\infty$ case where the SW expression \eqref{effective_T_K}
pierces the NRG data. This remains an interesting and experimentally
relevant open question. 

\section{Josephson current\label{sec:Josephson-current}}

For complete analysis of the three-terminal nanostructure we now address
the Josephson loop. The DC supercurrent is known to deviate from the
usual first Josephson law $I_{J}(\phi)=I_{c}\sin{\phi}$ nearby the
singlet-doublet transition. The quantum phase transition (or crossover)
reverses the Josephson current $I_{J}(\phi)$, and such a ``$0-\pi$
transition'' is due to the parity change of the induced electron pairing
$\langle\hat{d}_{\downarrow}\hat{d}_{\uparrow}\rangle$. It can be
practically achieved by tuning either the gate potential \cite{vanDam-2006,*Cleuziou-2006,*Jorgensen-2007,*Eichler-2009},
the superconducting phase \cite{Maurand-2012,*Delagrange-2015,*Delagrange-2016} or magnetic field \cite{Wentzell-2016}.

The charge flow from the superconducting $L$ electrode to the QD
can be derived using the Heisenberg equation of motion $I_{L}(\phi)=e\frac{d}{dt}\sum_{{\bf k},\sigma}\langle\hat{c}_{{\bf k}\sigma L}^{\dagger}\hat{c}_{{\bf k}\sigma L}\rangle=\frac{e}{i\hbar}\sum_{{\bf k},\sigma}\big\langle[\hat{c}_{{\bf k}\sigma L}^{\dagger}\hat{c}_{{\bf k}\sigma L},\hat{H}]\big\rangle$.
It can be formally expressed by the anomalous Green's function \cite{Zonda-2015,*Zonda-2016}
\begin{equation}
I_{L}(\phi)=\frac{2e}{\hbar\beta}\sum_{\omega_{n}}\frac{\Gamma_{S}\Delta}{\sqrt{\Delta^{2}+\omega_{n}^{2}}}\Im\left[\mb{G}_{12}(i\omega_{n})e^{i\phi/2}\right],\label{Josephson_current_formula}
\end{equation}
where $\omega_{n}=(2n+1)\pi\beta^{-1}$ are the fermionic Matsubara
frequencies. We have previously shown \cite{Zonda-2015,*Zonda-2016}
that \eqref{Josephson_current_formula} satisfies the charge conservation
$I_{L}(\phi)=-I_{R}(\phi)$ and is fully consistent with an alternative
derivation of the Josephson current from the phase derivative of the
free energy. In what follows we thus treat the DC Josephson current
as $I_{J}(\phi)\equiv I_{L}(\phi)$. Let us remark, that in the infinite-gap
limit $\Delta\rightarrow\infty$ Eq.\ \eqref{Josephson_current_formula}
simplifies to \cite{Oguri-2013} 
\begin{eqnarray}
I_{J}(\phi)=\frac{2e\Gamma_{S}}{\hbar}\;|\langle\hat{d}_{\downarrow}\hat{d}_{\uparrow}\rangle|\;\sin\left(\phi-\theta_{d}\right),\label{J_infinity}
\end{eqnarray}
where $\theta_{d}=0$ or $\pi$ is the phase of the induced pairing
$\langle\hat{d}_{\downarrow}\hat{d}_{\uparrow}\rangle=|\langle\hat{d}_{\downarrow}\hat{d}_{\uparrow}\rangle|\;e^{i\theta_{d}}$.
Expression \eqref{J_infinity} explicitly implies the Josephson
current reversal upon changing the parity $\theta_{d}$ between $0$
and $\pi$, which occurs at the singlet-doublet quantum phase transition
\cite{vanDam-2006,*Cleuziou-2006,*Jorgensen-2007,*Eichler-2009,Maurand-2012,*Delagrange-2015,*Delagrange-2016}.
\begin{figure*}[tp]
\includegraphics[width=1\textwidth]{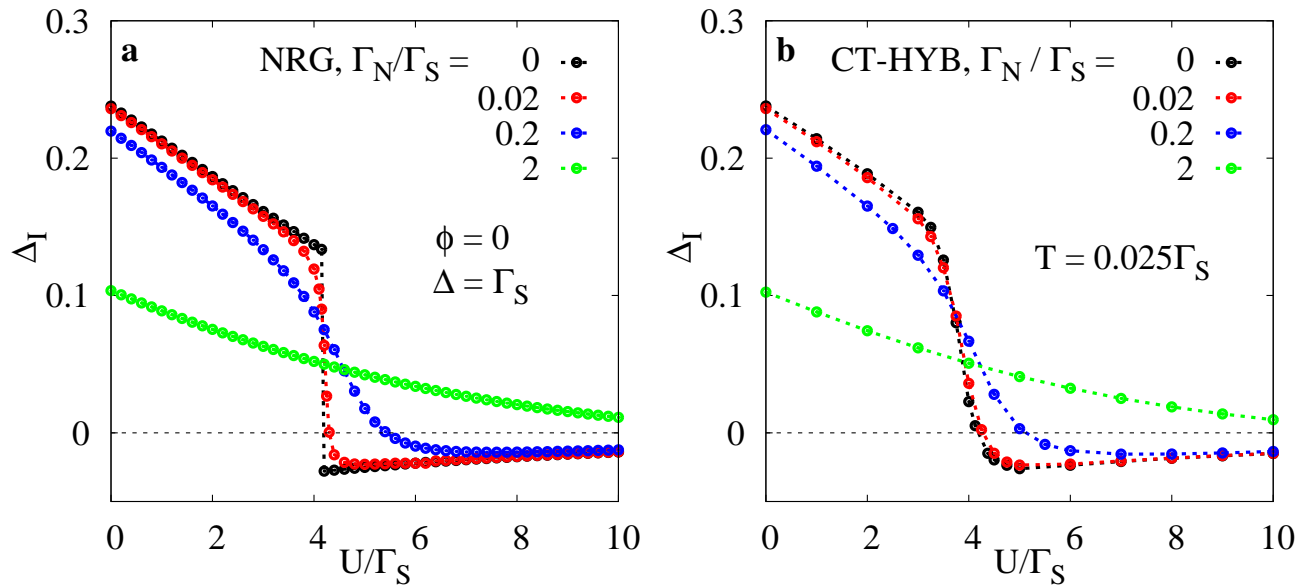} \caption{\label{fig:pairing_parameter} (Color online) Induced on-dot pairing
$\Delta_{I}=\langle\hat{d}_{\downarrow}\hat{d}_{\uparrow}\rangle$
as a function of the interaction $U_{d}$ calculated for finite gap
$\Delta=\Gamma_{S}$ and $\phi=0$ for several values of the normal
electrode coupling $\Gamma_{N}$. (\textbf{a}) NRG results for T=0.
(\textbf{b}) CT-HYB results for $T=0.025\Gamma_{S}$. QMC error bars
are smaller than the symbol size.}
 
\end{figure*}

In our three-terminal structure (Fig.\ \ref{fig:scheme}) the metallic
lead gives rise to a finite broadening of the subgap states, therefore
the singlet evolves to doublet (and {\em vice versa}) in a continuous
manner. This continuous crossover has already been discussed by Oguri
{\em et al.} \cite{Oguri-2013} for the superconducting atomic
limit $\Delta\rightarrow\infty$. Here we extend the approach to the
more realistic situations corresponding to finite SC gap $\Delta<\infty$
and nonzero temperature $T>0$.

Since the interplay of pairing and correlations affects the Josephson
current indirectly via the pair correlation $\Delta_{I}\equiv\langle\hat{d}_{\downarrow}\hat{d}_{\uparrow}\rangle$
we first consider its variation as a function of the interaction strength
$U_{d}$. Figure \ref{fig:pairing_parameter} shows the NRG (\textbf{a})
and CT-HYB (\textbf{b}) results obtained for representative couplings
$\Gamma_{N}$, as indicated. In the absence of the normal lead ($\Gamma_{N}=0$)
the induced pairing $\Delta_{I}$ changes abruptly at the quantum
phase transition where the parity $\theta_{d}$ evolves from $0$
to $\pi$. Under such circumstances the $0-\pi$ transition in S-QD-S
junctions is manifested by the sharp sign change of $I_{J}(\phi)$
\cite{vanDam-2006,*Cleuziou-2006,*Jorgensen-2007,*Eichler-2009,Maurand-2012,*Delagrange-2015,*Delagrange-2016}
(theoretically a discontinuity at zero temperature $T=0$). For either
finite couplings $\Gamma_{N}$ or finite temperature $T$ the induced
pairing $\Delta_{I}$ continuously changes from the positive to negative
values. Let us remark that the previous study \cite{Oguri-2013}
of the infinite-gap limit $\Delta\to\infty$ has not shown the negative
values of $\Delta_{I}$. On the other hand, for sufficiently strong
couplings $\Gamma_{N}$ the broadening of Andreev/Shiba states is
so large that they are no longer distinguishable (spectroscopically
they are seen as one broad structure). For this reason the induced
pairing $\Delta_{I}$ monotonously decreases with respect to $U_{d}$
as seen in the behavior for the large value $\Gamma_{N}=2\Gamma_{S}$.

\begin{figure*}[htpb]
\includegraphics[width=1\textwidth]{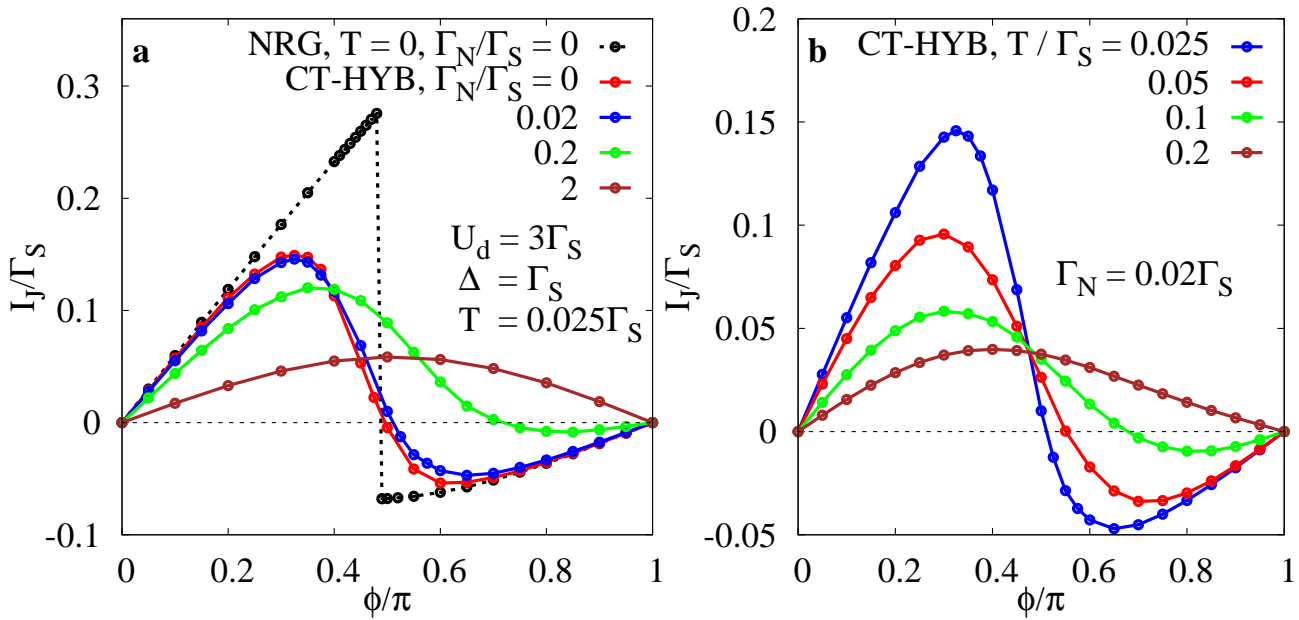} \caption{\label{fig:Josephson_current} (Color online) Phase dependence of
the DC Josephson current obtained from CT-HYB for $\Delta=\Gamma_{S}$
and $U_{d}=3\Gamma_{S}$ for fixed temperature $T=0.025\Gamma_{S}$
and changing $\Gamma_{N}$ (\textbf{a}) and fixed normal coupling
$\Gamma_{N}=0.02\Gamma_{S}$ and varying temperature (\textbf{b})
as indicated. In the left panel \textbf{a} we included for $\Gamma_{N}=0$
the zero-temperature NRG results for comparison. QMC error bars are
smaller than the symbol size. }
\end{figure*}

Practical observability of the effects discussed above can be done
by measuring the phase dependence of Josephson current. In Fig.\ \ref{fig:Josephson_current}
we plot the results (for $U=3\Gamma_{S}$ and $\Delta=\Gamma_{S}$)
of evaluating Eq.~\eqref{Josephson_current_formula} by CT-HYB analogously
to Refs.~\cite{Luitz-2010,*Luitz-2012}, since NRG is very inefficient
in this situation (finite $\Delta$ and nonzero $\phi$ and $\Gamma_{N}$
implying a three-channel problem, see above). The results shown in
the left panel \textbf{a} were obtained at fixed temperature $T=0.025\Gamma_{S}$.
We included the available zero-temperature NRG data for $\Gamma_{N}=0$.
Comparing the CT-HYB results with NRG shows that the chosen temperature
used in CT-HYB is low enough to provide a reasonable value of critical
$\phi_{c}$ for which the current changes sign. This point is pushed
to higher values of $\phi$ with increasing $\Gamma_{N}$ and for
high enough values of $\Gamma_{N}$ the region of the negative current
($\pi$ phase) is no longer present. Similar behavior is obtained
by changing the temperature for a fixed value of $\Gamma_{N}$ in
the right panel \textbf{b} \footnote{All the curves in Fig.~\eqref{fig:Josephson_current}\textbf{b} for
different temperatures seem to intersect at almost, but not exactly,
the same point. This behavior is consistent with other QMC results~\cite{Luitz-2010}.}. This indicates that indeed the normal coupling $\Gamma_{N}$ and
temperature have qualitatively the same effect on the system \textemdash{}
they both lead to the broadening of the Andreev/Shiba states and to
vanishing of the $\pi$ phase. 

\begin{figure}
\includegraphics[width=1\columnwidth]{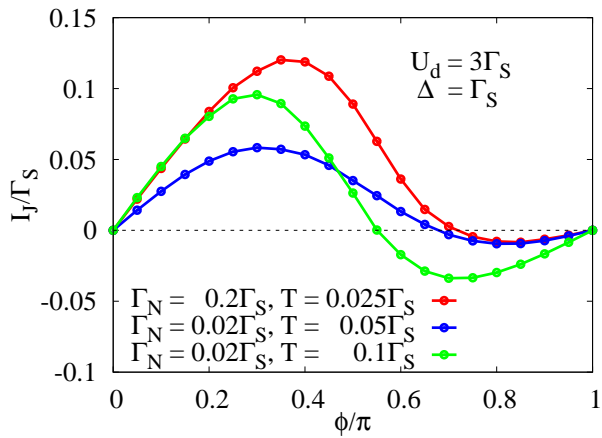}\caption{\label{fig:JC_comparison} (Color online) Comparison between the effects
of finite normal-lead coupling $\Gamma_{N}$ and temperature $T$.
Failure to fit the finite-$\Gamma_{N}$ (red) curve with finite-temperature
curve(s) (green \& blue) demonstrates the nonexistence of a quantitative
effective-temperature description of the normal lead influence (at
least) on the Josephson transport. }
\end{figure}

However, this qualitative statement cannot be extended to quantitative
predictions as we demonstrate in Fig.~\ref{fig:JC_comparison},
where we try to model a finite $\Gamma_{N}$ by an effective temperature.
More precisely, we attempt to mutually match two curves: one with
very small $T$ and finite $\Gamma_{N}$ (red line) and the other
with very small $\Gamma_{N}$ and finite effective temperature $T_{\mathrm{eff}}$
(blue and green curves). As can be seen from the figure, this attempt
fails \textemdash{} we can fit either the small- or high-$\phi$ parts
of the current-phase relations with various effective temperatures,
but we cannot find a single effective temperature which would cover
the whole phase range. Thus, the correspondence between the finite
normal-lead coupling $\Gamma_{N}$ and temperature is only a vague
qualitative analogy which should be used with much care. 

Unfortunately, the phase-dependent Josephson current does not indicate
directly any features that could be strictly assigned to the Kondo
effect. We hence think that the Josephson loop of the three-terminal
structure would be useful merely for spotting the singlet-doublet
crossover region, where the Andreev differential conductance should
reveal the characteristic zero-bias enhancement. Charge transport
through the Andreev loop would then allow for precise measurement
of the phase-tunable Kondo temperature as we now demonstrate.

\section{Andreev conductance\label{sec:Andreev-conductance}}

Empirical evaluation of the phase-controlled Kondo temperature can
be carried out by the charge transport in the N-QD-S branch of our setup. 
Under general nonequilibrium conditions the current is given by 
the lesser Green's function which should be determined from the Keldysh 
technique. However, for deeply subgap voltages $|V|\ll\Delta/e$ (of interest for our study of the subgap Kondo effect) the current can 
be expressed by the  Landauer-B\"uttiker formula \cite{Fazio-1998,*Krawiec-2004}
\begin{eqnarray}
I_{A}(V)=\frac{2e}{h}\int\!\!d\omega\;T_{A}(\omega)\left[f(\omega\!-\!eV)\!-\!f(\omega\!+\!eV)\right].\label{I_A}
\end{eqnarray}
This anomalous current is transmitted via the Andreev 
mechanism in which electrons from the conducting (N) reservoir are 
scattered into holes back to the same electrode, producing the Cooper 
pairs in superconducting (S) lead with the transmittance 
$T_{A}(\omega)\equiv\Gamma_{N}^{2}\left|\mb{G}_{12}(\omega+i0^{+})\right|^{2}$
which can be regarded as a quantitative measure of the induced on-dot pairing. Its maximal
unitary value is achieved for the BCS-type ground state configurations
when $eV$ coincides with the subgap Andreev/Shiba quasiparticle energies
\cite{Baranski-2013}. Nevertheless, also for the (spinful) doublet
configuration there is some contribution from the Andreev transport
at least nearby the doublet-singlet crossover \cite{Domanski-2016}.
We suggest this particular feature for probing the phase-dependent
Kondo temperature.

\begin{figure}
\includegraphics[width=1\columnwidth]{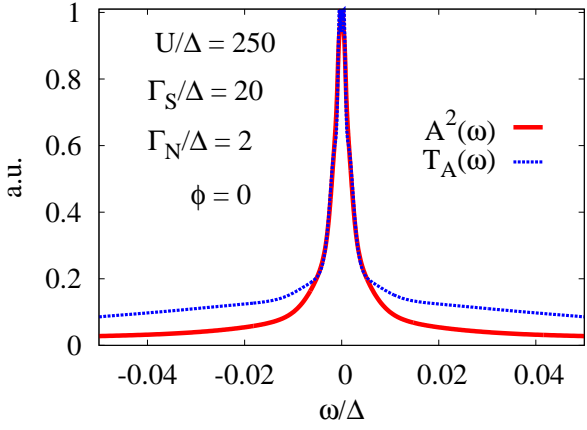}\caption{\label{fig:Andreev_comparison}(Color online) Comparison between the
square of the diagonal spectral function $A(\omega)$ of Sec.~\ref{sec:Spectral-function}
and the Andreev transmittance $T_{A}(\omega)$ showing their identical
zero-frequency-peak width which can be related to the Kondo scale.}
\end{figure}

To justify this proposal we first demonstrate that both the Andreev
transmittance $T_{A}(\omega)$ and the diagonal spectral function
$A(\omega)$ studied in detail in Sec.~\ref{sec:Spectral-function}
are governed by the very same low-energy scale, i.e., the Kondo scale
$T_{K}$. This means that the measurable width of the differential
Andreev conductance $G_{A}(V)\equiv dI_{A}(V)/dV$ peak at $V=0$
can be related to the Kondo scale $T_{K}$ theoretically studied in
Sec.~\ref{sec:Spectral-function} and, therefore, our theoretical
predictions can be experimentally verified. At sufficiently low temperature
the Andreev conductance $G_{A}(V)$ in terms of the conductance quantum
$2e^{2}/h$ is equal to twice the Andreev transmittance, $G_{A}(V)=4e^{2}/h\,T_{A}(eV)$,
and we thus just consider the two quantities interchangeably. 

Figure \ref{fig:Andreev_comparison} shows the (normalized) Andreev
transmittance $T_{A}(\omega)$ and square of the spectral function
$A(\omega)$ calculated by the NRG for a given set of parameters corresponding
to the $\pi$ phase with a well-developed Kondo peak. We can clearly
see that around $\omega=0$ both quantities overlap, which proves
that the zero-frequency (Kondo) peaks are determined by a single scale
which can be easily related to the Kondo scale studied in Sec.~\ref{sec:Spectral-function}.
Technically, the Andreev transmittance is given as the square of the
off-diagonal component of Green's function \eqref{GF} and that is
the reason why it should be compared with the square of the diagonal
spectral function which is determined by the diagonal part of Green's
function \eqref{GF}. This just means that the whole (matrix) Green's
function is governed by the single low-energy scale emerging as a
complex zero point of the Green's function determinant, cf.~Refs.~\cite{Zonda-2015,*Zonda-2016}. 

\begin{figure}[t]
\includegraphics[width=1\columnwidth]{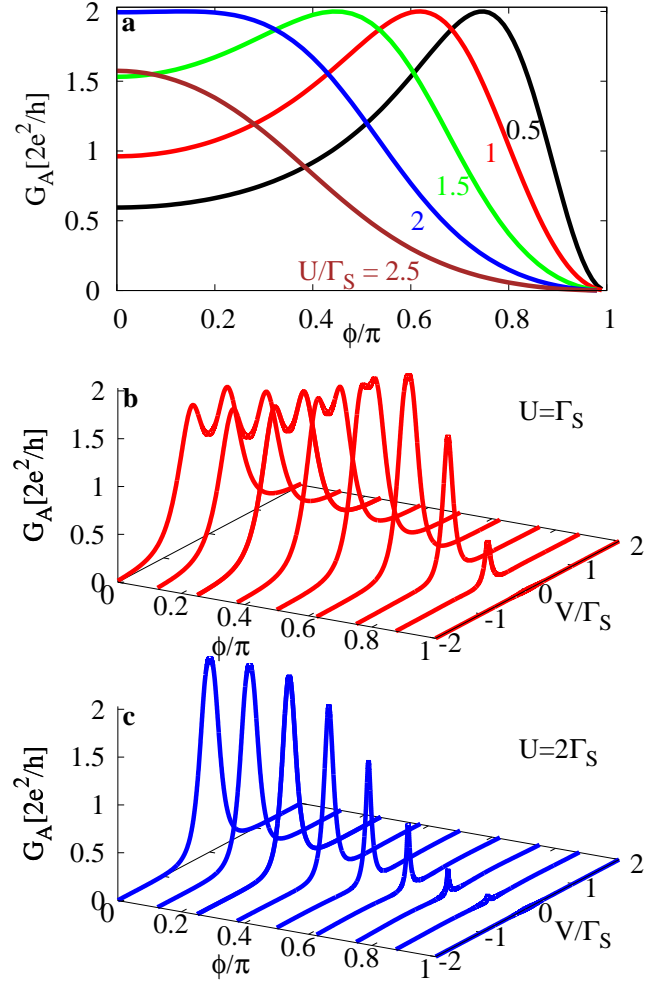} \caption{\label{fig:Andreev_conductance} (Color online) (\textbf{a}) Phase
dependence of the zero-bias Andreev conductance $G_{A}(V=0)$ in the
infinite-gap limit $\Delta\to\infty$ for $\Gamma_{N}/\Gamma_{S}=0.5$
and several values of the Coulomb potential $U_{d}$ as indicated.
(\textbf{b}) and (\textbf{c}) Differential Andreev conductance $G_{A}(V)$
as a function of the normal lead potential $V$ in the subgap regime
corresponding to two values of the Coulomb interaction $U_{d}$ from
the top panel \textbf{a}. All the results were obtained by NRG calculations.}
 
\end{figure}

In Fig.\ \ref{fig:Andreev_conductance} we plot the phase dependence
of the linear Andreev conductance $G_{A}(V=0)$ (panel \textbf{a})
as well as the finite-bias differential Andreev conductance $G_{A}(V)$
corresponding to two values of the Coulomb interaction $U_{d}=\Gamma_{S}$
(panel \textbf{b}) and $U_{d}=2\Gamma_{S}$ (panel \textbf{c}) from
the panel \textbf{a}. Thus the corresponding curves in panel \textbf{a}
are given as cuts $V=0$ of the graphs in the lower panels. Panel
\textbf{b} depicts a situation with the $0-\pi$ crossover realized
for an intermediate value of $\phi_{c}$ while in panel \textbf{c} the
system always stays in the $\pi$ phase regardless of the value of
$\phi\in(0,\pi)$. In the first situation (panel \textbf{b}), at small
phase difference $\phi$ the system is in the BCS-type (singlet) configuration,
we hence observe two maxima with the maximal conductance $G_{A}(V)=4e^{2}/h$.
They appear at such voltage $eV$, which corresponds to the quasiparticle
(Andreev/Shiba states) energies. Upon increasing the phase $\phi$
these maxima gradually merge at the critical phase $\phi_{c}$ where
the linear Andreev conductance reaches its maximal value $4e^{2}/h$.
At such a crossing point the ground state changes its configuration
and the system enters the (spinful) doublet state \cite{Oguri-2013}.
In that state the induced on-dot pairing is no longer efficient and,
therefore, the Andreev conductance is suppressed. In the vicinity
of such a singlet-doublet transition, from the doublet side, there
appears the characteristic phase-dependent zero-bias enhancement due
to the Kondo effect. The lower panel \textbf{c} for a stronger interacting
case does not exhibit any phase crossover, instead, we can observe
with decreasing phase $\phi$ the steady growth of the zero-bias peak,
whose width reveals the Kondo temperature. 

Actually, the experimental data obtained by Chang \emph{et al.} \cite{Chang-2013}
for an InAs nanowire confined between two superconducting Al leads and
the third normal Au electrode have already indicated such zero-bias
enhancement as a function of the back-gate potential, see the bottom
panel in their Fig.\ 2. We hope that experimental estimation of the
zero-bias width in a generalized SQUID setup enabling tuning of the
phase difference \cite{Cleuziou-2006,Maurand-2012,Delagrange-2015,Delagrange-2016}
will be soon feasible and our theoretical predictions will be compared
to the experimental data.

\section{Summary\label{sec:Summary}}

We have investigated the correlated quantum dot embedded in the Josephson
junction and additionally coupled to the normal metallic electrode.
Focusing on the subgap regime we have addressed an interplay between
the correlations and the induced on-dot pairing, emphasizing that
it can be controlled by the Josephson phase.

The tunable singlet-doublet crossover \cite{Oguri-2013,Paaske-2015}
has already been observed experimentally \cite{Delagrange-2015,Delagrange-2016}.
In this work we have proposed a feasible procedure for measuring the
phase-dependent Kondo temperature. In the subgap regime such Kondo
effect is driven by the effective spin-exchange between the QD and
normal lead electrons which operates only in the spinfull (doublet)
configuration. We have determined the phase-dependent Kondo temperature
$T_{K}$ by three independent methods, using (i) the numerical renormalization
group calculation, (ii) the self-consistent second order perturbation
treatment of the Coulomb potential, and (iii) the generalized Schrieffer-Wolff canonical transformation
projecting out the hybridization between the QD and normal lead electrons.
The last method implies the universal phase scaling $\ln{T_{K}}\propto\cos^{2}{\phi/2}$,
in agreement with the SOPT (at weak coupling) and the NRG (for arbitrary
$U_{d}$) results.

Charge currents induced in the Josephson and Andreev circuits both
reveal the phase-mediated crossover from the singlet to doublet configurations.
We have shown by QMC calculations that the usual $0-\pi$ transition
of the Josephson spectroscopy is strongly affected by the normal electrode:
For small (but finite) coupling $\Gamma_{N}$ the transition is continuous
whereas for larger couplings the $\pi$ phase is completely absent.
We have also shown that the nonlinear conductance of the Andreev current
can provide the detailed information on the subgap quasiparticle energies
and the Kondo temperature. Such differential conductance reveals the
zero-bias enhancement driven by the Kondo effect upon approaching
the doublet-singlet crossover. Width of this feature yields the characteristic
Kondo scale. We hope that our proposal will stimulate experimental
efforts for measuring the subgap Kondo temperature that can be fully
controlled by the Josephson phase.
\begin{acknowledgments}
This work is supported by the National Science Centre (Poland) through
the Grant No.\ DEC-2014/13/B/ST3/04451 (T.D., M.\v{Z}., T.N.), the Czech Science
Foundation via Projects No.\ 16-19640S (T.N.) and 15-14259S (V.P., V.J.),
and the Faculty of Mathematics and Natural Sciences of the University
of Rzesz\'{o}w through the project WMP/GD-12/2016 (G.G.). Access to computing
and storage facilities owned by parties and projects contributing
to the National Grid Infrastructure MetaCentrum, provided under the
programme ``Projects of Large Research, Development, and Innovations
Infrastructures'' (CESNET LM2015042), is greatly appreciated.
\end{acknowledgments}

%\bibliographystyle{apsrev4-1}
%\bibliography{Josephson}
%\include{Final.bbl}

%

\end{document}